\documentclass[twocolumn,aps,showpacs]{revtex4}
\usepackage[dvips]{graphics,color}
\usepackage{graphicx}
\usepackage{dcolumn}
\usepackage{bm}
\usepackage{epsfig}

\newcommand{\beq}{\begin{equation}}
\newcommand{\eeq}{\end{equation}}
\newcommand{\beqa}{\begin{eqnarray}}
\newcommand{\eeqa}{\end{eqnarray}}
\newcommand{\bd}[1]{ \mbox{\boldmath $#1$}}

\begin{document}
\def\ii{\'\i}

\title{
Extending the GKZ limit without breaking Lorentz Invariance
}

\author{P. O. Hess$^{1,2}$ and Walter Greiner$^{1}$}
\affiliation{
$^1$Frankfurt Institute of Advanced Studies, Johann Wolfgang Goethe Universit\"at,
Max-von-Laue Str. 1, 60438 Frankfurt am Main, Germany \\
$^2$Instituto de Ciencias Nucleares, UNAM, Circuito Exterior, C.U., A.P. 70-543, 04510 M\'exico D.F., Mexico
}

\begin{abstract}
{
A scenario is presented on how to shift the predicted cutoff
in the cosmic ray spectrum at $10^{20}$~eV,
called the Greisen-Zatsepin-Kuzmin limit (GKZ), to larger energies
without breaking the Lorentz invariance. The formulation is based on
a pseudo-complex extension of standard field theory.
The dispersion relation of particles can be changed, leading
to a modification of the GKZ limit. Maximal shifts are determined.
}
\end{abstract}

\maketitle


The GZK limit
(Greisen-Zatsepin-Kuzmin) \cite{gkz1,gkz2}
corresponds to a cutoff in high energy cosmic rays at
$10^{20}$~eV.
Experiments were performed \cite{fly,haverah,yakutsk,agasa}
to measure the spectrum of these energy cosmic rays.
In AGASA \cite{agasa} several events breaking this limit
were observed. The new
experiment of Auger \cite{auger} in Argentina intends to verify these data.
First results are reported in Ref. \cite{lukas}.
If confirmed, it will be generally considered
as a clear evidence for the breaking of Lorentz symmetry.

The GKZ limit can be calculated considering a head on collision of a
proton with a photon from the Cosmic Microwave back ground (CMB),
producing a pion, the lightest hadron produced in such a collision.
The standard dispersion relation $E=\sqrt{p^2+m^2}$
is assumed. Thus, a change in the
GZK limit is associated with a change of this dispersion relation.
The main stream of explanations is to model
violation of Lorentz invariance.

One particular example for breaking Lorentz invariance is presented
in \cite{amelia,dsr2}, denominated {\it Double Special
Relativity}. Besides the constant $c$
a smallest fundamental length scale ($L$) is introduced.
Any length is treated as a vector and is subject to a contraction.
In order to keep the smallest scale invariant, the boost
transformation has to be deformed, allowing contractions
only for lengths larger than $L$.
This requires a deformation of
the boost, implying an explicit breaking of Lorentz invariance.
The procedure is phenomenologic.

An alternative way is to {\it introduce a larger group} than the
Lorentz one,
which leads to an extended field theory. We show here 
a possible scenario in which
the Lorentz invariance
may be maintained, while the GKZ limit is lifted to larger
energies.
It relies on an extended field theory, which has been proposed in Refs.
\cite{schuller1,schuller2}, requiring a pseudo-complexification of
space.
The pseudo-imaginary component of the generators of the Poincar\'e
group are related to transformations to accelerated/rotated systems.
A maximal acceleration appears, which is inverse to a
fundamental length scale $l$ (we adopt the natural units where
$\hbar = c = 1$).
The length scale is
{\it not subject to a Lorentz contraction} but is a {\it scalar}
as the velocity of light is.
We also predict a maximal shift of the GKZ limit, depending on particular
phenomenological considerations. Though, a derivation from first principles
is recommended, this is the best we can do for the moment.
Nevertheless, a phenomenologic approach is often the best way to gain a first
insight into new physics.
The main point
is that there is an open possibility not to break Lorentz invariance!
Work for a complete formulation of the theory is in progress.

The pseudo-complexification of space-time requires the introduction
of pseudo-complex numbers (also known as hypercomplex). We will first enlist
some of their their main properties, necessary for the understanding of the
present contribution. For the mathematical details, please consult
Refs. \cite{crumeyrolle,schuller1}.


The pseudo-complex numbers are {\it defined} via
$X = x_1 + I x_2$ = $x_+ \sigma_+ + x_- \sigma_-$,
with $I^2=1$ and $\sigma_\pm = \frac{1}{2} \left( 1 \pm I \right)$.
This is similar to the complex notation except for
the different behavior of $I$.
The $\sigma_\pm$ obey the relations $\sigma_\pm^2=\sigma_\pm$ and
$\sigma_+\sigma_-=0$. The last expression of $X$ above is
in terms of the {\it zero divisor basis} (a linear combination in terms
of $\sigma_\pm$), which turns out to
be extremely useful.

The pseudo-complex conjugate of a pseudo-complex number is 
$X^*  =  x_1 - I x_2 = x_+\sigma_- + x_- \sigma_+$.
The {\it norm} of a pseudo-complex number is given by
the square root of
$|X|^2 = XX^ *  =  x_1^2 - x_2^2$.
There are three different possibilities: $|X|^2>0$, $|X|^2<0$ and $|X|^2>0$.
The structure of this space is isomorphic to $O(1,1)$.

Calculations, like
differentiation and integration, can be done in complete analogy to the
case of complex numbers.
The pseudo-complex derivative is denoted by $\frac{D}{DX}$ and
the rules of application are the same.
The derivative
can be directly extended to variables with an additional index
($\frac{D}{DX_\mu}$) and
to functional derivatives.

However, no residual theorem exists.
Thus, the structure of pseudo-complex
numbers is very similar to the usual complex numbers but not completely,
due to the appearance of the zero divisor branch
${\cal P}^0={\cal P}^0_+ \cup {\cal P}^0_-$, with
${\cal P}^0_\pm = \left\{\lambda\sigma_\pm | \lambda \epsilon \bd{R} \right\}$.
It implies that the set of pseudo-complex numbers is just a {\it ring}.
This reflects the less stringent algebraic structure.


In the pseudo-complex Lorentz group $SO_{\bd{P}}(1,3)$,
finite transformations are given by
$exp(i\omega_{\mu\nu}M^{\mu\nu})$, with $\omega_{\mu\nu}$
a pseudo-complex number and $M^{\mu\nu}$ the generators of the
Lorentz group. The usual Lorentz group is obtained by
restricting to real numbers.

The particular combination
$M_\pm^{\mu\nu}$  =  $\frac{1}{2}\left(M^{\mu\nu}\pm I M^{\mu\nu} \right)$,
leads to commuting generators $M_+^{\mu\nu}$ with $M_-^{\mu\nu}$.
This implies that the pseudo-complex Lie group is the direct product of
two algebras, each with the commutation relations of a $SO(1,3)$ group.
In group notation we write

\beqa
SO_{\bd P} (1,3) \simeq SO_+(1,3) \otimes SO_-(1,3) \supset SO(1,3) ~~~.
\label{ps-sym}
\eeqa
This is the larger group we were looking for.

The pseudo-complex Poincar\'e group is generated by the pseudo-complex
four momentum

\beqa
P^\mu & = & iD^\mu = i \frac{D}{DX_\mu}= P^\mu_+ \sigma_+ + P^\mu_-\sigma_-
~~~,
\eeqa
and generators of the pseudo-complex Lorentz algebra.

A Casimir operator of the pseudo-complex Poincar\'e group is

\beqa
P^2 & = \sigma_+ P^2_+ + \sigma_- P^2_- ~~~.
\eeqa
Its eigenvalue is $M^2 = \sigma_+ M_+^2 + \sigma_- M_-^2$, i.e.,
a pseudo-complex mass associated to each particle.
Similarly, the Pauli-Ljubanski vector \cite{gen-lo} can be defined.

In this new approach, fields are direct extensions from standard field
theory: If a field transforms under a given representation
of the Lorentz group, the pseudo-complex extension of the field transforms
in the same way in both components ($\sigma_+$ and $\sigma_-$). For example,
a Weyl-spinor transforms as $(\frac{1}{2},0)+(0,\frac{1}{2})$ under
the Lorentz group. The same transformation property holds in the pseudo-complex
field with respect to the $SO_+(1,3)$ and $SO_-(1,3)$ groups.
One consequence is that in a field $\Psi$ =
$\Psi_+ \sigma_+ + \Psi_- \sigma_-$ {\it both components} ($\Psi_+$
and $\Psi_-$) {\it have the same spin}.


A new variational principle is introduced \cite{schuller2},
in order to connect both zero divisor components,
i.e. $\delta S ~\epsilon~ {\cal P}^0$,
a number in the zero divisor branch.
If this condition would not be imposed, the result is two equations
of motion of two independent system, i.e., no new field theory
would be obtained.

When physical observables are calculated, a projection
to the pseudo-real part is applied. E.g., $P_2^\mu$ the pseudo-real part of the
linear momentum, related to acceleration, is set to zero.
It corresponds to go into an inertial system, where the
vacuum is well defined.
This is interpreted as going back to the inertial system, which is
connected to an accelerated system via a pseudo-imaginary transformation.

For the case of a Dirac field, the Lagrange density is given by
${\cal L} = \bar{\Psi} \left(\gamma_\mu P^\mu - M \right)\Psi$ and
the resulting field equation, after variation, is

\beqa
(\gamma_\mu P^\mu - M) ~\epsilon~ {\cal P}^0 ~~~,
\eeqa
with ${\cal P}^0$ being the set of zero divisors,
as defined in the introduction.
Multiplying with the pseudo-conjugate $(\gamma_\mu P^\mu - M)^*$,
leads to

\beqa
(\gamma_\nu P_+^\nu -M_+)(\gamma_\mu P_-^\mu -M_-^2)\Psi & = & 0 ~~~,
\label{p+p-}
\eeqa
which is the final field equation for free fields.

Now, inspecting Eq. (\ref{p+p-}), a solution is given either by setting the
first or the second factor equal to zero, having substituted before
$P^\mu_\pm$ by $p^\mu$. Without loss of generality, let us take the first choice.
Then, the solution describes a propagating particle with mass $M_+=m$,
which is identified as the physical mass. The other mass parameter,
$M_-$ is related to the regulating mass of the theory, of the order of
$\frac{1}{l}$. Due to the appearance of $M_-$ the theory
is regularized a la Pauli-Villars \cite{schuller3},
which is an important detail.

In this theory, the photon is described
with $M_+=0$ but $M_-=\frac{1}{l}$, a large regularizing mass.
This generates a term in the Lagrange density of the form
$\sim M_-^2A_\mu A^\mu \sigma_-$, which assures gauge invariance
because of the appearance of $\sigma_-$.
This is due to the property that the gauge angle $\alpha (x)$
has the same value
in the $\sigma_+$ and $\sigma_-$ component, thus, 
$\alpha (x) \sim \sigma_+$, but $\sigma_+$ and $\sigma_-$ commute.

In conclusion, the advantages of the pseudo-complex field theory are:
i) It is regularized, ii) stays gauge invariant
and, the most important point,
iii) it maintains known symmetries and thus permits to
proceed in a very similar fashion to the standard field theory and the
determination of cross sections for different processes.


Having resumed the most important characteristics of the extended
field theory, we proceed to the main part.
Our proposal is to add to the Lagrange density an interaction
of the type $\bar{\Psi}lI\gamma^\mu f_\mu \Psi$, which is a scalar under the
Lorentz group $SO(1,3)$, diagonally embedded in
$SO_+(1,3) \otimes SO_-(1,3)$.

A possible reason for this term is as follows:
It is led by the assumption that during a collision
new {\it effective interactions} result from
contributions of accelerated systems, connected to the inertial in- and
outgoing systems.
Because there is a minimal length scale, the interaction is distributed
over a finite size of space-time, which is a consequence of a finite,
maximal acceleration, not permitting an instant interaction.
A transformation to an accelerated system is
given by $exp(lI\omega \cdot L)$
= $exp(l\omega \cdot L)\sigma_+ + exp(-l\omega \cdot L)\sigma_-$,
with $L_i$ being a generator of the
Poincar\'e group. The smallness of $l$ indicates the order of contribution.
Applying it to $\Psi (x)$, expanding up to first order,
yields (note that $I=\sigma_+-\sigma_-$ and define $\bd{\mu}$
= $exp(l\omega \cdot L)$)

\beqa
& \frac{\left( \Psi (\bd{\mu}x) + \Psi (\bd{\mu}^{-1}x) \right)}{2}
+ lI\frac{\left( \Psi (\bd{\mu}x) - \Psi (\bd{\mu}^{-1}x) \right)}{2l} &
\nonumber \\
& \approx \Psi (x) + lI\overrightarrow{\bd{\Pi}}\Psi (x) ~~~,
\label{change}
\eeqa
where the first part is an average over field values from
neighboring systems, and the second term gives the
contribution to the field, due to the transformation to
accelerated/rotated systems.
It describes the {\it differences of fields} to neighboring accelerated systems.
The action of this transformation
is symbolically expressed by the operator $\bd{\Pi}$ on
the state vector. The arrow indicates the direction of action.
In principle, one should model the interaction by weighting over the
contributions of different accelerated systems.
The exact form depends on the transformation parameters and their range,
i.e., on the model used.

Considering now the term $\bar{\Psi}\gamma_\mu p^\mu \Psi$, with the usual
linear momentum $p_\mu$. Changing $\Psi$ by the expression in
(\ref{change}), leads in zero order in $l$ to the old term and a correction
$\sim lI$ of first order in $l$:

\beqa
\bar{\Psi}\gamma_\mu p^\mu \Psi & \rightarrow
\bar{\Psi} \left( p^\mu + I P_2^\mu \right)\Psi
+ lI \bar{\Psi} \gamma_\mu \left( \overleftarrow{\bd{\Pi}}p^\mu
+ p^\mu \vec{\bd{\Pi}} \right) \Psi  ~~~,
\eeqa
where $IP_2^\mu$ comes from the pseudo-complexification of the linear
momentum.
As an alternative,
the last term can be modeled as $\bar{\Psi} \gamma_\mu lIf^\mu \Psi$.
Because of dimensional reasons the $f^\mu$ has the
dimensions of a force which leads to a particular model.
{\it A more strict derivation is needed, but for the moment this
phenomenologic argument suffices to present our point}.

Instead of this heuristic explanation, we can skip it and
start from the ad hoc addition of the term
just discussed.

The advantage lies in the possibility
to include this within a generalization of the {\it minimal coupling
scheme}.
The extended minimal coupling, concerning the linear momentum, reads

\beqa
P_\mu \rightarrow P_\mu + lIf_\mu ~~~,
\label{extension}
\eeqa
where the $l$ is due to dimensional reasons and
indicates that the correction should be of the order
of the minimal length scale.
After having projected to an inertial system ($P_2^\mu = 0$),
the $f_\mu$ is still present and describes the role of a non-vanishing,
pseudo-imaginary component of the linear momentum, reflecting the
effect of the acceleration.

Let us suppose that $f_\mu$ is given by

\beqa
f_\mu & \sim & {\cal F}_{\mu\nu}p^\nu ~~~,
\label{model1}
\eeqa
with ${\cal F}_{\mu\nu}$ being an antisymmetric tensor not defined yet.
This choice of $f_\mu$
contains the possibility to interpret $f_\mu$ as a force
($f_\mu = \frac{dp_\mu}{d\tau}$, $\tau$ being the eigen-time, thus,
$p_\mu f^\mu = 0$).
Then, under the modification (\ref{extension})
the gauge transformation for the electro-magnetic vector field
$A_\mu$ \cite{schuller3} is also changed adding to the transformation a term
$\frac{l}{g}If_\mu \alpha (x)$, with $\alpha (x)$ as the gauge angle and
$g$ the coupling strength to the electro-magnetic field.

This is not the only possible choice. The $f_\mu$ may
in general depend on a certain power $\beta$ in the linear momentum,
i.e. symbolically

\beqa
|f_\mu | & \sim & p^\beta ~~~.
\eeqa
The effects, as a function in $\beta$ will be discussed further below
for two cases.

This leaves  a rich choice of possible interactions which can be
invented, studying their effects on the dispersion relation.

The resulting equation of motion for $\Psi$,
setting $A_\mu$ to zero, is modified to

\beqa
\left( \gamma^\mu (P_\mu + lIf_\mu) - M \right) \Psi ~~~\epsilon ~~~
{\cal P}^0 ~~~,
\eeqa
with ${\cal P}^0$ = ${\cal P}_+^0 \cup {\cal P}_-^0$ (see the definition
given above), is the set of zero divisors.
Multiplying by the pseudo-complex conjugate of the operator,
multiplying by $(\gamma_\mu \left[ P^\mu_- - lf_\mu \right] + M_-)$
$(\gamma_\mu \left[ P^\mu_+  + lf_\mu \right]+ M_+)$ and
using the properties of the $\gamma^\mu$ matrices,
we arrive at the equation

\beqa
\left( {\cal P}_{+\mu} {\cal P}^{\mu}_+ - M_+^2 \right)
\left( {\cal P}_{-\mu} {\cal P}^{\mu}_- - M_-^2 \right) & = & 0 ~~~.
\eeqa
We project to an inertial system ($P_2^\mu = 0$), i.e.
${\cal P}^\mu_\pm=p^\mu \pm lf^\mu$.
Selecting the first factor, using
$P_{+\mu} P^{\mu}_+= E^2-p^2+l^2f_\mu f^\mu$
$+$ $l(p_\mu f^\mu + f_\mu p^\mu)$, we arrive at the
dispersion relation

\beqa
E^2 = p^2 + (lf)^2 +l(pf+fp)+M_+^2 ~~~,
\eeqa
with $f^2=-f_\mu f^\mu >0$.
and $pf=-p_\mu f^\mu$, $fp=-f_\mu p^\mu$.
Interpreting $f_\mu$ as a force, i.e. a derivative of $p_\mu$ with respect
to the eigen-time, we have $p_\mu f^\mu =0$, which eliminates
the term proportional to the first order in the 
length scale.

Let us now investigate the consequences for the threshold momentum
for the production of pions, without specifying $f_\mu$.
Using energy and momentum conservation in a head-on collision,
it was shown in \cite{dsr1} that the threshold linear momentum of a
reaction $B_1+\gamma \rightarrow B_2 + M_3$ ($B_k$ being baryons,
e.g. protons, and $M_3$ being a meson, e.g. pion, and $\gamma$ is
a soft photon from the CMB) is given by

\beqa
p_{1,{\rm thr.}} & \approx & \frac{(m_2+m_3)^2-m_1^2}{4\omega}
~=~10^{11}~GeV ~~~,
\label{head-on}
\eeqa
with $\omega$ being the energy of a soft photon from the CMB
and $m_k$ the masses of the participating particles.

With the modified dispersion relation, the energy changes to
($pf=fp=0$)

\beqa
E_k & \approx & p_k + \frac{m_k^2}{2p_k} + \frac{l^2f_k^2}{2p_k} ~
 = ~ p_k + \frac{\tilde{m}_k^2}{2p_k} ~~~,
\label{approx2}
\eeqa
with $\tilde{m}_k^2 = m_k^2 + l^2 f_k^2$ and $f_k^2 = -f_{k\mu}f_k^\mu$ 
for particle number $k$ = (1,2,3).
The result for the threshold momentum is now

\beqa
p_{1,{\rm thr.}} & \approx & \frac{(\tilde{m}_2+\tilde{m}_3)^2-
\tilde{m}_1^2}{4\omega}
~ \approx ~ \frac{(m_2+m_3)^2-m_1^2}{4\omega}  \nonumber \\
& & +\frac{l^2}{4\omega} \left[ (m_2+m_3)
\left( \frac{f_2^2}{m_2} + \frac{f_3^2}{m_3} \right) -f_1^2 \right] ~~~.
\label{head-on2}
\eeqa
The new threshold momentum is shifted to larger values if
the expression in the square bracket is positive. Similar
terms are added, when we skip the above constriction of
$pf=fp=0$.

{\bf We can estimate how far the threshold momentum can be shifted}:
As one example, assume that the last term in Eq. (\ref{head-on2}) is
proportional to $p^2$ ($\beta = 2$),
which corresponds to the first choice of $f_\mu$ above.
We arrive at a quadratic equation
$p-Xp^2 = Y$, where $Y\approx 10^{11}$~GeV.
The dependence on $l$ is hidden in $X$.
The solution
is $p_{1/2}  =  \frac{1}{2X} \left( 1 \pm \sqrt{1-4XY} \right)$.

\begin{figure}[t]
\centerline{\epsfxsize=7cm\epsffile{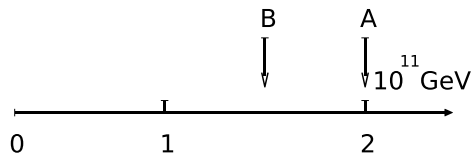}}
\vskip -5cm
\caption{
Schematic illustration of the shift of the GKZ to larger values. {\bf A}
refers to
the ansatz of $f_\mu$ as a force, proportional to the linear momentum $p$ and
{\bf B} refers to the estimation of $f_\mu$ as an average force.
}
\label{fig1}
\end{figure}

For each given $X$ we have, thus, two solutions. Taking the minus sign,
we reproduce in the limit of $X\rightarrow 0$ the solution $p=Y$.
The largest value we obtain for $X=\frac{1}{4Y}$ and the threshold
momentum is $2*10^{11}$~GeV, i.e., {\bf just the double value}.
We stress, that this result is only valid when $f^2$ is proportional to
$p^2$. It changes when the dependence in the power of $p$ is different.
The maximal shift by a factor of 2 is the consequence of the second
order equation in $p$. The exact value of the shift depends on
$X\sim \frac{l^2}{\omega}$, which may serve to determine $l$.

The $f_\mu$ may be approximated by
the average force acting on the particle
($\beta = 0$).
The expression in Eq. (\ref{head-on2}) stays the same, with the difference
that it adds some constant to the right hand side of the equation,
implying a shift of the threshold momentum to higher values.
Assuming $f^2_k=(m_k a)^2$, $f^2_3<<f^2_1=f^2_2$ and the acceleration "$a$" as
maximal, i.e. $\frac{1}{l}$, leads 
{\bf only to a shift to 1.5~10$^{11}$~GeV}.
When the terms proportional to $pf$ and $fp$ are included,
the change involves also a rescaling of $p_{1,thr}$, which corresponds
to  the case of $\beta = 1$. However, it implies a different interpretation
than $f_\mu$ being a force.

Which form of $f_\mu$ finally describes the physical situation
depends on several assumptions, which may change the present results.
The explanation given here is of
phenomenological 
nature and not much can be said about the exact structure of $f_\mu$. 
It would be more attractive to get $f_\mu$ from basic principles.
For that, we have to complete the formulation of
the pseudo-complex field theory.
Nevertheless, phenomenologic considerations shed some light on possible
processes. Work is still in progress.

The main point given here is that
the GZK limit can be shifted to larger values
without assuming a violation of Lorentz invariance,
if one accepts to change
standard field theory. The pseudo-complex field theory seems to be one
candidate. It has several advantages, which we list here again:
The theory is regularized, maintains gauge invariance
and it keeps the known symmetries.
This facilitates the calculation of cross sections, following standard
procedures.

Note, that most models explaining a breaking of Lorentz invariance,
like given in \cite{dsr2}, are phenomenologic in nature. 
Considering the lack of a fundamental theory, this is the only viable way
up to now. The pseudo-complex extension of field theory seems to be
the correct direction to go.



\vskip 1cm

\section*{Acknowledgments}

Financial support from the DFG, DGAPA and CONACyT. P.O.H.
is a Mercator-professor and thanks the FIAS
for all the support and hospitality given.
We acknowledge very helpful comments by F. P. Schuller from
the ICN-UNAM.

\end{document}